# Utilizing RNA-Seq Data for Cancer Network Inference

Ying Cai[1,2], Bernard Fendler[1], Gurinder S. Atwal[1]
1. Cold Spring Harbor Laboratory, Quantitative Biology, Cold Spring Harbor, NY, USA
2. Applied Mathematics and Statistics, Stony Brook University, Stony Brook, NY, USA

*Abstract*—An important challenge in cancer systems biology is to uncover the complex network of interactions between genes (tumor suppressor genes and oncogenes) implicated in cancer. Next generation sequencing provides unparalleled ability to probe the expression levels of the entire set of cancer genes and their transcript isoforms. However, there are onerous statistical and computational issues in interpreting high-dimensional sequencing data and inferring the underlying genetic network. In this study, we analyzed RNA-Seq data from lymphoblastoid cell lines derived from a population of 69 human individuals and implemented a probabilistic framework to construct biologically-relevant genetic networks. In particular, we employed a graphical lasso analysis, motivated by considerations of the maximum entropy formalism, to estimate the sparse inverse covariance matrix of RNA-Seq data. Reverse engineering the network of genetic isoforms revealed a layer of genetic regulatory complexity not exhibited by traditional microarrays. Gene ontology, pathway enrichment and protein-protein path length analysis were all carried out to validate the biological context of the predicted network of interacting cancer gene isoforms.

*Keywords*— RNA-Seq; graphical lasso; maximum entropy; cancer

## I.  INTRODUCTION

High-throughput sequencing has become an important alternative to microarray assays, with RNA-Seq gaining in recent popularity. RNA-Seq is more sensitive and has a larger dynamic range [1] than traditional microarrays. Moreover, RNA-Seq has low background noise, which enables it to provide more information to detect allele-specific expression and identify novel promoters, isoforms, novel exons and splice sites [2]. As such, the development of RNA-Seq has become an important tool to delineate the full complexity of transcription across the genome.

Tumor suppressor genes (TSGs) and oncogenes form a coordinated network of genes that respond to a wide variety of genotoxic stresses in normal cells and tumors. These stress responses are orchestrated via controlled levels of gene transcription, and it remains a fundamental challenge in cancer systems biology to uncover the entire network of transcriptional regulation. We propose to utilize RNA-Seq data to infer the cancer transcriptional network, and in doing so we address several machine learning challenges that are incurred when dealing with sparse noisy high-dimensional data.

A variety of statistical approaches have been applied to reconstruct transcriptome networks, such as information-theoretic methods [3] and relevance networks methods [4]. However, these reverse-engineering methods all focus on quantifying the statistical correlations, as opposed to the direct interactions, between each pair of genes in order to build a genetic network. Inspired by concepts in statistical physics, gene interactions were inferred by a different study through the application of the maximum entropy method using microarray expression training data [5]. This approach illustrated that, under general assumptions, the elements of the *inverse* covariance matrix provide an appropriate measure of pairwise gene interaction, in contradistinction to correlation-based methods that focus on just the covariance matrix itself. We propose to utilize and build upon this formalism with respect to the analysis of recent RNA-Seq data in the hopes of uncovering a principled understanding of the underlying cancer transcriptome network.

In genomic studies, the number of genes is typically much larger than the number of samples, resulting in an undersampled and noninvertible covariance matrix. To prevent overfitting we can provide an estimate of the sparse inverse covariance matrix using the graphical lasso framework [6]. The graphical lasso algorithm uses L1 regularization to control the number of zeros in the inverse covariance matrix in order to learn the structure in an undirected Gaussian graphical model.

In this study, RNA-Seq data of 474 cancer-related genes (known TSGs and oncogenes) and their corresponding isoforms were obtained from 69 human cell line samples. The processed sequencing dataset was then analyzed to infer and understand the network of pairwise interactions amongst all the genes. By working with RNA-Seq data, as opposed to microarray expression values, we were also able to reverse engineer the network of interactions between differing gene isoforms, revealing an extra layer of genetic regulatory complexity. To provide biological validation of the resulting predicted networks we carried out Gene Ontology (GO), pathway enrichment and protein-protein path length analyses.

## II. MATERIALS AND METHODS

### A. Data Set Pre-Processing and Correction

RNA-Seq data were obtained from a recent study whereby lymphoblastoid cell lines derived from a sample of 69 Yoruban (African Hapmap population) individuals were sequenced on the Illumina GAII platform (details described in [2]). Approximately 1.2 billion short reads were generated and mapped to the hg18 reference genome using TopHat (v1.3.3) [7]. The mapping results were then further analyzed using Cufflinks (v1.1.0) [8] with default parameters. The expression level of each gene was defined using the FPKM (fragment per kilobase of exon per million fragments mapped) values as reported by Cufflinks.

Gene/isoform expression level is a noisy phenotype and a wide range of external factors, such as batch effects, can confound its measurement. We employed the recent Bayesian PEER framework [9] to uncover the global effects of known and hidden factors on gene/isoform expression level. After estimating the confounding covariates in RNA-Seq expression data, which we found to be surprisingly significant, we then corrected the gene/isoform expression levels to obtain modified point estimates of expression.

In this study, we initially focused on a catalogue of 474 known tumor suppressor genes (TSGs) [10-12] and oncogenes [11, 13, 14]. Genes/isoforms with expression equal to zero across all samples were removed, leaving 417 cancer genes, and 1014 isoforms.

### B. Entropy Maximization

What is the most general probabilistic network model that captures the pairwise dependencies between genes/isoforms? The maximum entropy framework provides a principled answer to this question. Let the vector $\vec{x} = (x_1,...x_P)$ denote the expression levels of the P genes/isoforms. Let $p(\vec{x})$ denote the probability distribution function, and N the total number of samples. The most general and unstructured probability distribution that captures the pairwise correlations is the one that maximizes the entropy $H = -\sum_{\vec{x}} p(\vec{x}) \ln p(\vec{x})$ [5], subject to the normalization and expectation constraints: $\sum_{\vec{x}} p(\vec{x}) = 1$,
$\langle x_i \rangle = \sum_{\vec{x}} p(\vec{x}) x_i = \frac{1}{N} \sum_{k=1}^{N} x_i^k$,
$\langle x_i x_j \rangle = \sum_{\vec{x}} p(\vec{x}) x_i x_j = \frac{1}{N} \sum_{k=1}^{N} x_i^k x_j^k$.

The maximization over the entropy functional is carried out by introducing Lagrange multipliers ($\alpha, \beta$, and $M$),

$$J = H - \alpha \sum_{\vec{x}} p(\vec{x}) - \sum_{i=1}^{P} \beta_i \sum_{\vec{x}} p(\vec{x}) x_i - 2 \sum_{i,j=1}^{P} M_{ij} \sum_{\vec{x}} p(\vec{x}) x_i x_j.$$

Setting $\frac{\delta J}{\delta p(\vec{x})} = 0$, we have a Boltzmann-like distribution $p(\vec{x}) \propto e^{-S}$, where $S = \frac{1}{2} \sum_{ij} y_i M_{ij} y_j$ and $y_i = x_i + \sum_j M_{ij}^{-1} \beta_j$. $M$ specifies the pairwise interaction terms and, in the continuum limit, the correlation function is then simply calculated to be the inverse of the elements of $M$, $M_{ij}^{-1} = \langle y_i y_j \rangle = \langle x_i x_j \rangle - \langle x_i \rangle \langle x_j \rangle = C_{ij}$. Thus the inverse covariance matrix provides a principled measure of pairwise interaction between genes. In the language of statistical physics, the inverse covariance matrix represents the energy coupling constants between the variables $x_i$ and $x_j$.

### C. Graphical Lasso

As discussed in the Introduction, the covariance matrix is undersampled and thus is noninvertible. The graphical lasso technique obviates this issue, sparsifying the inverse covariance matrix $M$ and preventing over-fitting. This approach assumes that the FPKM observations of RNA-Seq data follow a multivariate Gaussian distribution. Formally, the inverse covariance matrix $M$ is estimated by maximizing the penalized log-likelihood [6],

$$\log \det M - \text{tr}(QM) - \rho \|M\|_1$$

where we have defined $M = C^{-1}$ and $Q$ is the empirical covariance matrix of the data. $\rho$ is a nonnegative tuning parameter. $\|M\|_1$ is the $L_1$ norm, the sum of the absolute values of the elements of $M$. When $\rho$ is sufficiently large, the estimate $\widehat{M}$ will be sparse due to the lasso-type penalty on the elements of $M$. In this study, we performed graphical lasso analysis for a range of tuning parameters, and here we report the results for $\rho = 0.01$, representing a suitable tradeoff between sparsity and running time of the calculations. Separate analyses were performed on both i.) genes (represented by the most common gene isoforms) and ii.) the entire set of isoforms across all genes. Gene/isoform pairs with the most significant interaction coupling constants (top 86 pairs among studied gene pairs, top 43 pairs among studied isoform pairs) were selected to represent the underlying cancer transcriptome network.

### D. Gene Ontology (GO) and Pathway Enrichment Analysis

Fisher's exact test was performed to determine which GO terms are significantly overrepresented in the set of the strongest interacting gene pairs in relation to the GO background of the total 417 cancer genes. In our study, overrepresented GO categories in biological process with p-values less than 0.05 were considered significantly enriched.

Kyoto Encyclopedia of Genes and Genomes (KEGG) is an online database integrating genomic, chemical, and systemic function information [15]. Fisher's exact test with a threshold of 0.05 was performed to determine whether the top interacting genes are significantly enriched in bimolecular pathways as compared to the background of the total 417 cancer genes.

### E. Protein-Protein Interactions

To examine if the gene pairs we identified were more functionally relevant than background cancer genes, we examined the protein-protein interactions for the gene pairs using a protein-protein interaction database, Human Protein Reference Database (HPRD) [16]. The functional relevance of a gene pair can be evaluated by examining the shortest path length between the two genes in the protein-protein interaction network [17]. It is reasonable to expect that gene pairs with higher interaction energy are expected to have shorter paths between them in the protein-protein interaction network. To make comparisons with correlations based methods we also calculated the shortest protein-protein distance between highly correlated gene pairs. The null distribution of network distances was calculated from random sets of cancer genes.

## III. RESULTS

### A. RNA-Seq Analysis and Graphical Lasso

The most significant interaction terms were then used to construct genetic networks and visualized using Cytoscape version 2.8 [18]. Figure 1 represents the gene network with the 20 strongest interaction terms and Figure 2 shows the isoform network with the strongest 15 isoform pairs. Interestingly, the isoform network did not closely match the gene network, indicating that most common isoforms do not represent the most informative elements of the transcriptional network amongst cancer-related genes.

### B. Gene Ontology and Pathway Enrichment

Several GO categories were considered as significantly enriched among the interacting genes with Fisher's exact p-value less than 0.05, compared to the background GO of the entire set of cancer genes. In total, 21 categories were recognized and the 7 most significant among them are shown in Table 1.

Pathway enrichment analysis based on the KEGG pathway database was carried out in this study. Significantly enriched pathways with Fisher's exact p-value less than 0.05 are listed in Table 2.

### C. Path Length Analysis

76 out of the 86 unique gene pairs with high interaction couplings selected from the sparse inverse covariance matrix were found to be present in the protein interaction database, HPRD. We carried out a Monte Carlo permutation analysis, comprising of 10,000 sets of 76 random gene pairs from the 417 cancer genes and computed their mean shortest path lengths. The null distribution for mean shortest path lengths was computed using a histogram of these values (see Figure 3). We found that the mean shortest path length (3.14) was significantly shorter for the top pairs (p = 0.03) than the mean of the null distribution. To compare this result with networks inferred from traditional correlation-based approaches the same procedure was applied to gene pairs selected from the covariance matrix. The null distribution for mean shortest path lengths is shown in Figure 4. Nevertheless, the mean shortest path length (3.23) was not significantly shorter for the top pairs (p = 0.17) than the mean of the null distribution. This result was consistent with the prediction that the strongest interacting gene pairs selected using our methodology are more likely to be functionally related with each other compared to the same number of top gene pairs selected from elements of the covariance matrix.

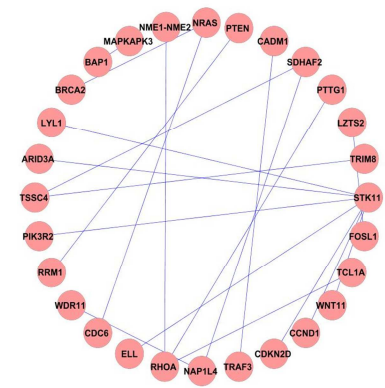

Figure 1. Inferred network of the strongest 20 pairwise cancer gene interactions. Each gene was represented by its most common isoform.

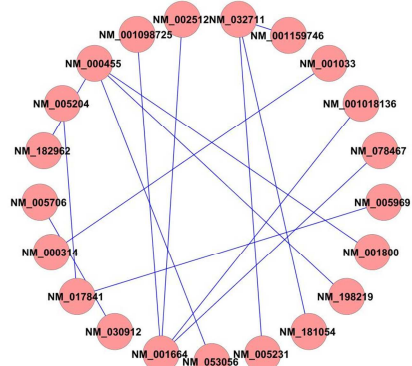

Figure 2. Inferred network of the strongest 15 pairwise cancer gene isoform interactions.

TABLE 1. GENE ONTOLOGY ENRICHMENT RESULTS

| Term | Genes | $-\log_{10}$(P-Value) |
|---|---|---|
| GO:0050890 cognition | NRAS, HRAS, JUN, FOSL1, MYC, PTEN, AXIN1 | 2.62 |
| GO:0007611 learning or memory | NRAS, HRAS, JUN, FOSL1, PTEN | 2.17 |
| GO:0022402 cell cycle process | CDC6, LZTS2, STK11, BRCA2, SMAD3, AURKA, PTTG1, HMGA2, CCND1, CDKN2D, GFI1, MYC, PINX1 | 1.96 |
| GO:0009628 response to abiotic stimulus | NRAS, HRAS, CCND1, JUN, CDKN2D, BCL3, BRCA2, GFI1, FOSL1, MYC | 1.92 |
| GO:0045596 negative regulation of cell differentiation | CCND1, LMO2, NME1-NME2, RHOA, TCL1A, SMAD3, GFI1 | 1.89 |
| GO:0050877 neurological system process | NRAS, HRAS, JUN, FOSL1, MYC, PTEN, AXIN1 | 1.89 |
| GO:0016055 Wnt receptor signaling pathway | CCND1, LZTS2, FRAT2, WNT11, AXIN1 | 1.82 |

TABLE 2. KEGG PATHWAY ENRICHMENT RESULTS

| Term | Genes | $-\log_{10}$(P-Value) |
|---|---|---|
| hsa04310:Wnt signaling pathway | CCND1, JUN, RHOA, SMAD3, FRAT2, WNT11, FOSL1, MYC, RBX1, AXIN1 | 3.70 |
| hsa04530:Tight junction | NRAS, HRAS, RHOA, YES1, PTEN | 1.74 |
| hsa05213:Endometrial cancer | NRAS, HRAS, CCND1, MYC, PTEN, AXIN1, PIK3R2 | 1.60 |
| hsa04350:TGF-beta signaling pathway | RBL2, RHOA, SMAD3, MYC, RBX1 | 1.57 |

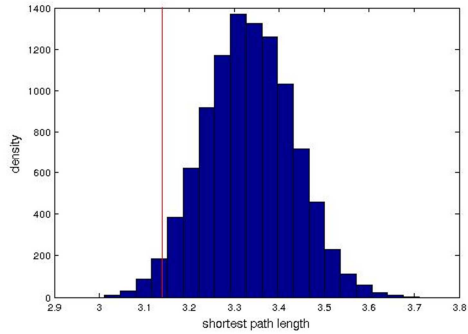

Figure 3. Null Monte Carlo distribution of shortest path lengths between gene pairs. The vertical line indicates the mean value of shortest path lengths (3.14) for the top interacting pairs selected from the sparse *inverse* covariance matrix. p-value 0.03.

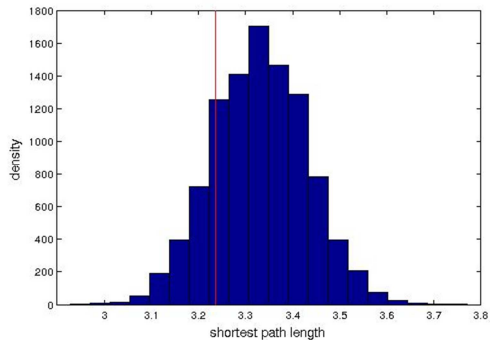

Figure 4. Null Monte Carlo distribution of shortest path lengths between gene pairs. The vertical line indicates the mean value of shortest path lengths (3.23) for the top interacting pairs selected from the covariance matrix. p-value 0.17.

## IV. DISCUSSION

Our study provides a comprehensive insight into the cancer gene/isoform transcriptome network in lymphoblastoid cell lines using RNA-Seq data, a powerful next generation DNA sequencing platform. RNA-Seq is an emerging effective method for transcriptome analysis and provides significantly higher levels of transcript accuracy than traditional microarray technology.

From the GO analysis results, we found many interacting genes related to cell cycle processes, indicating that the regulation of cell division requires strong cooperation between cancer-related genes. Pathway enrichment results yielded significant pathways including the Wnt signaling and the TGF-beta signaling pathway, indicating that these pathways need to be more tightly regulated than other cancer pathways. The results for the mean protein-protein path length between the predicted interacting genes lend further weight to the biological relevance of the gene pairs. Together, all these results demonstrate that the sparse inverse covariance method, as motivated by the maximum entropy formalism, provides a powerful framework to uncover biologically relevant network models of RNA-Seq cancer data, based on pairwise interactions. Our future work will apply the statistical framework described in this paper to the whole genome, and extend the analysis to higher-order interactions among genes.

## ACKNOWLEDGEMENTS

This work was supported by the Simons Foundation [Y.C, B.F. and G.S.A] and the Starr Cancer Consortium, I3-A123 [G.S.A.].